# Gravitational polarization of the quantum vacuum caused by two point-like bodies

Dragan Slavkov Hajdukovic and Sergej Walter
INFI, Cetinje, Montenegro
dragan.hajdukovic@cern.ch

## Abstract

In a recent paper (Hajdukovic 2020a) quantum vacuum was considered as a source of gravity and, the simplest phenomenon, the gravitational polarization of the quantum vacuum by an immersed point-like body, was studied. In the present paper, we have derived *the effective gravitational charge density* of the quantum vacuum, caused by two immersed point-like bodies. Among others, the obtained result proves that quantum vacuum can have regions with a *negative* effective gravitational charge density. Hence, quantum vacuum, the "ocean" in which all matter of the Universe is immersed, acts as a complex fluid with a very variable gravitational charge density that might include both positive and *negative densities*; a crucial prediction that can be tested within the Solar System. In the general case of $N \geq 3$ point-like bodies, immersed in the quantum vacuum, the analytical solutions are not possible, and the use of numerical methods is inevitable. The key point is that an appropriate numerical method, for the calculation of the effective gravitational charge density of the quantum vacuum induced by N immersed bodies, might be crucial in description of galaxies, without the involvement of dark matter or a modification of gravity. The development of such a valuable numerical method, is not possible, without a *previous* (and in this study achieved) *understanding of the impact of a two-body system*.

## 1. Introduction

Recently the gravitational field of a point-like body immersed in the quantum vacuum was studied (Hajdukovic 2020a) under the working hypothesis that *quantum vacuum fluctuations are virtual gravitational dipoles*. It is well-established that quantum vacuum contains extremely short-living virtual electric dipoles (for instance electron-positron and quark-antiquark pairs which are in fact two electric charges of the opposite sign). The simplest mental picture is to consider a virtual gravitational dipole as an analogue of a virtual electric dipole, i.e., two gravitational charges of the opposite sign. However, there is a fundamental difference between electric and gravitational charges; the gravitational charges of the same sign attract each other, while charges of the opposite sign repel each other.

If correct, the hypothesis that quantum vacuum fluctuations are virtual gravitational dipoles is the simplest solution to the cosmological constant problem making possible to consider quantum vacuum as a source of gravity; a so far forgotten source of gravity.

The key point is that the otherwise randomly oriented virtual gravitational dipoles, are aligned (to some extent – completely or only partially) with the external gravitational field caused by matter immersed in the quantum vacuum. In a region of non-random orientation of virtual gravitational dipoles, the gravitational polarization density $\boldsymbol{P}_g$, i.e., the gravitational dipole moment per unit volume, can be attributed to the quantum vacuum. It is obvious that the magnitude $|\boldsymbol{P}_g|$ of the gravitational polarization density $\boldsymbol{P}_g$ satisfies the inequality $0 \leq |\boldsymbol{P}_g| \leq P_{gmax}$ where 0 corresponds to the random orientations of dipoles, while the maximal magnitude $P_{gmax}$ corresponds to the case of saturation (when all dipoles are aligned with the external field).

While the gravitational charge density of the quantum vacuum is everywhere zero, an *effective* gravitational charge density that acts as a real one (Hajdukovic 2020a) can be attributed to the quantum vacuum:

$$\rho_{qv} = -\nabla \cdot \mathbf{P}_g. \tag{1}$$

Let us underscore that, from purely mathematical point of view the effective gravitational charge density can be both positive and *negative*; positive when $\nabla \cdot \mathbf{P}_g < 0$ and negative when $\nabla \cdot \mathbf{P}_g > 0$. As we will demonstrate analytically at the end of Section 2 and numerically in Section 3, already in the case of gravitational polarization caused by two bodies, there are regions in which the effective gravitational charge density (calculated using Eq. (1)) has negative value.

The effective gravitational charge density of the quantum vacuum depends on the distribution of the immersed matter. Two simplest, but fundamental cases, are the gravitational polarization caused by a point-like body and the gravitational polarization caused by two point-like bodies that will be studied in the present paper. Let us underscore that only in these two cases, the effective gravitational charge density of the quantum vacuum can be expressed analytically as a function of masses and coordinates; *for three or more point-like bodies the use of numerical methods is inevitable*.

The understanding of the gravitational polarization of the quantum vacuum that is caused by two point-like bodies is crucial for the eventual development of an accurate numerical method for the general case of $N \geq 3$ bodies. Hence, the present study can be considered as a first step towards *a new kind of N-body simulations that include the gravitational impact of the polarized quantum vacuum*. Among others, if the gravitational polarization of the quantum vacuum exists in nature, such simulations might describe galaxies (Hajdukovic 2011, Hajdukovic 2014) without the involvement of dark matter or a modification of gravity.

## 2. The effective gravitational charge density caused by two point-like bodies

In general, the magnitude $|\mathbf{P}_g|$ is a function of the used coordinates $(x_1, x_2, x_3)$, while the gravitational polarization density $\mathbf{P}_g$ and the corresponding Newtonian acceleration $\mathbf{g}_N$ point in the same direction. Consequently (Hajdukovic 2020a), we can write:

$$|\mathbf{P}_g| = P_{g\max} f_g(x_1, x_2, x_3) \ and \ \mathbf{P}_g = P_{g\max} f_g(x_1, x_2, x_3) \frac{\mathbf{g}_N}{|\mathbf{g}_N|}. \tag{2}$$

The values of function $f_g(x_1, x_2, x_3)$ belong to the interval $[0,1]$ where values 0 and 1 correspond respectively to the cases of random orientation of dipoles and *saturation*. For simplicity, in the following text, the magnitude $|\mathbf{g}_N|$ will be often denoted by $g_N$.

Equations (1) and (2) lead to:

$$\rho_{qv}(x_1, x_2, x_3) = -P_{g\max} \nabla \cdot \left[ f_g(x_1, x_2, x_3) \frac{\mathbf{g}_N}{|\mathbf{g}_N|} \right]. \tag{3}$$

Hence, in order to calculate the effective gravitational charge density of the quantum vacuum (analytically for $N \leq 2$ and numerically for $N \geq 3$ point-like bodies) we must know the function $f_g(x_1, x_2, x_3)$ and the unit vector $\mathbf{g}_N/|\mathbf{g}_N|$ of the Newtonian acceleration caused by the considered system of bodies.

A single point-like body immersed in the quantum vacuum perturbs the random orientation of dipoles and creates around itself *a single spherical halo* of the polarized quantum vacuum. Because of spherical symmetry, it is natural to use spherical coordinates $(r, \theta, \varphi)$; the unit vector $\mathbf{g}_N/|\mathbf{g}_N| = -\mathbf{r}_0$, where $\mathbf{r}_0$ denotes the unit vector of the radial coordinate $r$. So, from Eq. (3), the effective gravitational charge density and the total gravitational charge of the quantum vacuum within a sphere of radius $r$ are given by:

$$\rho_{qv}(r) = P_{g\max} \frac{1}{r^2} \frac{d}{dr} [r^2 f_g(r)]. \tag{4}$$

$$M_{qv}(r) = 4\pi P_{g\max} r^2 f_g(r). \tag{5}$$

A reasonable approximation for $f_g(r)$ is (Hajdukovic 2020a):

$$f_g(r) = \tanh\left(\frac{R_{\text{sat}}}{r}\right)\tanh\left(A\frac{R_{\text{sat}}}{r}\right); \quad R_{\text{sat}} = \sqrt{\frac{M_b}{4\pi P_{\text{gmax}}}}. \qquad (6)$$

In the above equation, $M_b$ is the baryonic mass of the body while A is a dimensionless constant equal to the ratio of the total effective gravitational charge of the Universe and the baryonic mass of the Universe in the time of the birth of the Cosmic Microwave Background radiation; hence, if we assume that what we call dark matter is the effective gravitational charge of the quantum vacuum, $A \approx 5.3$ (Hajdukovic 2020a) for an isolated *point-like* body. However, if a composite system is roughly considered as a point like body (for instance a galaxy at sufficiently large distance from its center) the value of $A$ depends on the internal structure and can significantly differ from $A \approx 5.3$.

Eq. (6) is a particular case for the single body (Hajdukovic 2020a), and already in the case of two point-masses (and multi-body systems in general) we must use its generic formulation:

$$f_g(x_1, x_2, x_3) \equiv f_g(g_N) = \tanh\left(\sqrt{\frac{g_N}{g_{\text{qvmax}}}}\right)\tanh\left(A\sqrt{\frac{g_N}{g_{\text{qvmax}}}}\right); \quad g_{\text{qvmax}} = 4\pi G P_{\text{gmax}}. \qquad (7)$$

So far, before a more accurate approximation is found, Eq. (7) must be used in the fundamental Eq. (3).

Now, let us focus on the case of two point-like bodies with masses $m_1$ and $m_2$ at mutual distance D. Because of cylindrical symmetry we will use the cylindrical coordinates (See Figure 1) usually denoted $(z, \rho, \varphi)$; however, in order to avoid confusion between density $\rho_{qv}$ and coordinate $\rho$, we will use the notation $(z, s, \varphi)$ and will denote the corresponding unit vectors $(\mathbf{e}_z, \mathbf{e}_s, \mathbf{e}_\varphi)$. As it can be easily deduced from Fig. 1, vector $\mathbf{g}_N = g_{Nz}\mathbf{e}_z + g_{Ns}\mathbf{e}_s$ can be explicitly written as:

$$\mathbf{g}_N = -\left\{\frac{Gm_1 z}{(z^2+s^2)^{3/2}} - \frac{Gm_2(D-z)}{[(D-z)^2+s^2]^{3/2}}\right\}\mathbf{e}_z - \left\{\frac{Gm_1 s}{(z^2+s^2)^{3/2}} + \frac{Gm_2 s}{[(D-z)^2+s^2]^{3/2}}\right\}\mathbf{e}_s. \qquad (8)$$

The corresponding magnitude of acceleration is:

$$|\mathbf{g}_N| \equiv g_N = G\sqrt{\frac{m_1^2}{(z^2+s^2)^2} + \frac{2m_1 m_2(z^2+s^2-Dz)}{(z^2+s^2)^{3/2}[(D-z)^2+s^2]^{3/2}} + \frac{m_2^2}{[(D-z)^2+s^2]^2}}. \qquad (9)$$

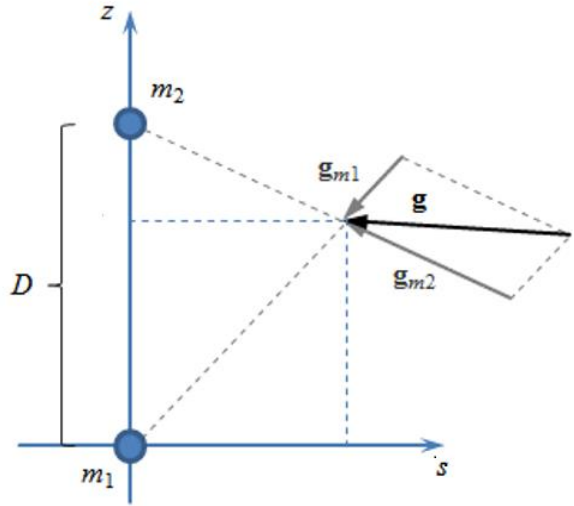

Figure 1. The resultant Newtonian gravitational field $g$ of two point-like bodies with mass $m_1$ and $m_2$ at mutual distance $D \equiv 2d$, has cylindrycal symmetry that trivially leads to Eq. (8).

For our two-body system with cylindrical symmetry, the basic equation (3) can be written as:

$$\rho_{\text{qv}}(z,s) = -P_{\text{gmax}}\left\{\frac{\partial}{\partial z}\left[f_g(z,s)\frac{g_{Nz}(z,s)}{g_N(z,s)}\right] + \frac{1}{s}\frac{\partial}{\partial s}\left[sf_g(z,s)\frac{g_{Ns}(z,s)}{g_N(z,s)}\right]\right\}. \qquad (10)$$

Equations (7), (8) and (9) determine all functions that appear on the right-hand side of Eq. (10); hence, the effective gravitational charge density of the quantum vacuum, as an explicit function of cylindrical coordinates, can be found by elementary but lengthy partial derivations over $z$ and $s$ (See Appendix A for details).

Let us note that the simplest case, approximation $f_g(g_N) \equiv f_g(z,s) = 1$ (i.e., the case of saturation) is of real importance. For instance, the Newtonian gravitational field caused by the Sun at a distance of 100 AU is $\approx 5.9 \times 10^{-7}\, m/s^2$; this is 4 orders of magnitude larger than the characteristic acceleration $g_{qvmax} < 6 \times 10^{-11}\, m/s^2$. Consequently, saturation is excellent approximation nearly everywhere in Solar System; important exception are regions in which gravitational fields of Sun and other celestial bodies cancel each other to values smaller than $g_{qvmax}$.

The next Section is completely devoted to numerical results based on equations (A1) to (A5) from the Appendix. However, before it, let us point out two general features of the explicit analytical solution.

The first important feature is the "granularity" in the distribution of the effective gravitational charge of the quantum vacuum. In the case of a single point-like body, there is a single halo of the polarized quantum vacuum having spherical symmetry and a finite size. In the case $N \geq 2$, the gravitational polarization of the quantum vacuum is a result of the common gravitational field of all bodies; however, individual halos of all bodies remain a major feature (each halo is limited to a region in which the gravitational field of the corresponding body is strongly dominant).

Second, in the case of a single body (Hajdukovic 2020a), the effective gravitational charge density of the surrounding quantum vacuum is always positive. An important question (See note after Eq. (1)) is if, in the case of systems composed of $N \geq 2$ bodies, the effective gravitational charge density of the quantum vacuum can also have negative values. The explicit function $\rho_{qv}(z,s)$ determined in the Appendix is too complicated to answer this question analytically but we can have some analytical insight considering the simplified case of the effective gravitational charge density caused by two equal masses $m$, in the plane $z = D/2 \equiv d$.

$$\rho_{qv}(z=d,s) = \frac{P_{gmax}}{s} f_g(z=d,s) \frac{2s^2 - d^2}{d^2 + s^2} - P_{gmax} F(z=d,s) m^2 s \frac{2s^2 - d^2}{(d^2+s^2)^4} =$$
$$= P_{gmax} \frac{2s^2-d^2}{d^2+s^2} \left[ \frac{f_g(z=d,s)}{s} - F(z=d,s) \frac{m^2 s}{(d^2+s^2)^3} \right] \qquad (11)$$

where $f_g(z=d,s)$ and $F(z=d,s)$ are functions that always have a positive value

$$f_g(z=d,s) = \tanh\left(\sqrt{\frac{g_N}{g_{qvmax}}}\right) \tanh\left(A\sqrt{\frac{g_N}{g_{qvmax}}}\right); \quad g_N \equiv g_N(z=d,s) = \frac{2Gms}{(d^2+s^2)^{3/2}} \qquad (12)$$

$$F(z=d,s) = \frac{2G^2}{g_N^{3/2}\sqrt{g_{qvmax}}} \left[ \frac{\tanh\left(A\sqrt{g_N/g_{qvmax}}\right)}{\cosh^2\left(\sqrt{g_N/g_{qvmax}}\right)} + \frac{A\tanh\left(\sqrt{g_N/g_{qvmax}}\right)}{\cosh^2\left(A\sqrt{g_N/g_{qvmax}}\right)} \right] \qquad (13)$$

We can liberate from hyperbolic functions in equations (12) and (13) if inequality $g_N \ll g_{qvmax}$ is valid at *all* points in the plane $(d,s,\varphi)$. If condition $g_N \ll g_{qvmax}$ is satisfied, hyperbolic functions can be well approximated by the first two terms in the corresponding Taylor series, $\tanh(x) \approx x - x^3/3$ and $1/\cosh(x) \approx 1 - x^2/2$). After such a simplification of right-hand sides of equations (12) and (13) and introduction of the obtained approximate values in Eq. (11), the result is:

$$\rho_{qv}(z=d,s) = \left[ \frac{4}{3}(A^3 + A) P_{gmax} \frac{G^2 m^2}{g_{qvmax}^2} \right] \frac{(2s^2-d^2)s}{(d^2+s^2)^4} \qquad (14)$$

Consequently, the sign of the effective gravitational charge density is determined by the sign of $2s^2 - d^2$; $\rho_{qv}(z = d, s)$ is negative for $0 < s < d/\sqrt{2}$ (See Figure 2). In brief, quantum vacuum can have regions with a negative gravitational charge density. Hence, from the gravitational point of view, quantum vacuum, the "ocean" in which all matter of the Universe is immersed, may act as a complex fluid with a very variable gravitational charge density that might include both positive and negative densities. The understanding of the impact of such a complex "fluid" in a galaxy (or in our Solar System) is not possible without the appropriate numerical methods which development is advocated in this paper.

The inevitable question is if the assumption $g_N \ll g_{qvmax}$ is an unacceptable oversimplification. Inequality $g_N \ll g_{qvmax}$ is *always* satisfied in the limits $s \to 0$ and $s \to +\infty$ (this is an obvious result from the second of equations (12)). However, inequality $g_N \ll g_{qvmax}$ can be also valid for *all* $s$ if the distance $D = 2d$ between bodies is sufficiently large so that $Gm/d^2 \ll g_{qvmax}$. For instance, this is true for our galaxy Milky Way and our neighbor Andromeda, while it is not necessarily true for a binary system of two supermassive black holes at a relatively small distance of one parsec. Hence, there are physical cases in which the condition $g_N \ll g_{qvmax}$ is valid for all $s$, but there are also systems in which this condition is wrong.

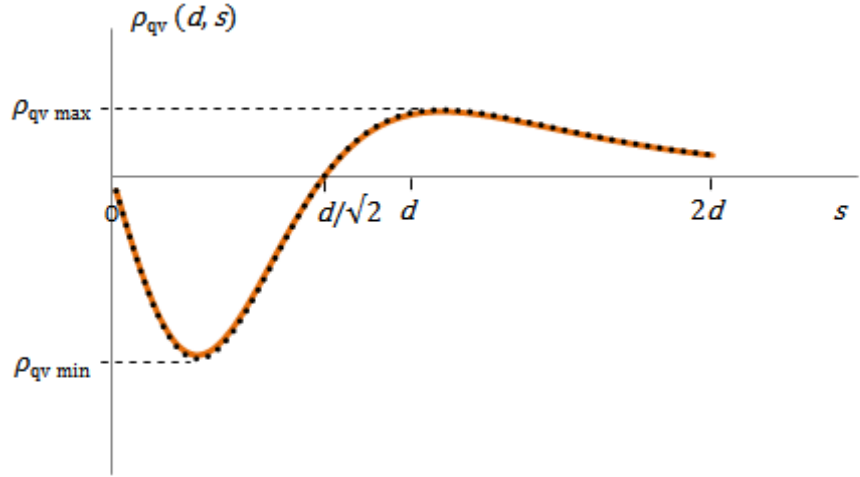

Figure 2. The effective gravitational charge density $\rho_{qv}(d, s)$, i.e. density in the plane $z = d$, between 2 bodies of the same mass $m$, at a distance $D \equiv 2d \gg R_{sat}$. Note the existence of interval $(0, d/\sqrt{2})$ in which density has a *negative* value; some regions of the quantum vacuum might act as if they have a *negative gravitational charge*. Red line and dached black line correspond respectively to exact function (Eq. (11)) and to its approximation (Eq. (14)).

## 3. Some illuminating numerical results for a two-body system

### 3.1 General features

In Appendix A, the exact effective gravitational charge density of the quantum vacuum is given as an exact function of masses of two bodies ($m_1$ and $m_2$), their mutual distance $D$ and cylindrical coordinates $(z, s, \varphi)$. Unfortunately, the analytical solution is so long that main features are practically hidden. However, it is quite easy to use the exact solution to create illuminating plots.

Before we continue let us clarify two points. First, all the plots are made on the z-s plane; consequently, the actual 3D volumes of the distribution of the effective gravitational charge densities are the solids of revolution of the plots around the z axis. Second, within a plot, densities differ at least a few orders of magnitude; hence the colour scale given in Fig. 4 is just a rough indicator.

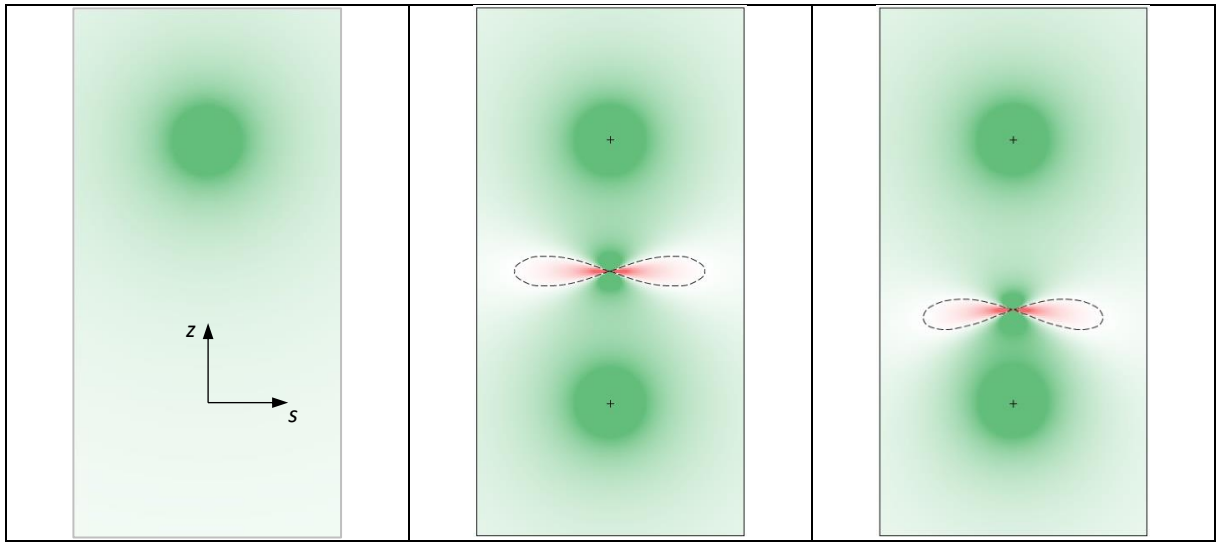

Figure 3. An illustrative distribution of gravitational charge density $\rho_{qv}(z,s)$ with mass of upper body arbitrary taken to be equal to mass of Sun. From left to right: the case of a single point-like body, the case of two bodies with the same mass, and the case of two bodies with different mass ( $m_1 = 0.3 m_2$ ). Dashed line denotes points with zero density; inside the region enclosed by dashed line density is *negative*, while outside that region density is positive. The green areas highlight positively charged gravitational density, *the red ones are for negative values* and at white spots density is negligible compared with green and red regions.

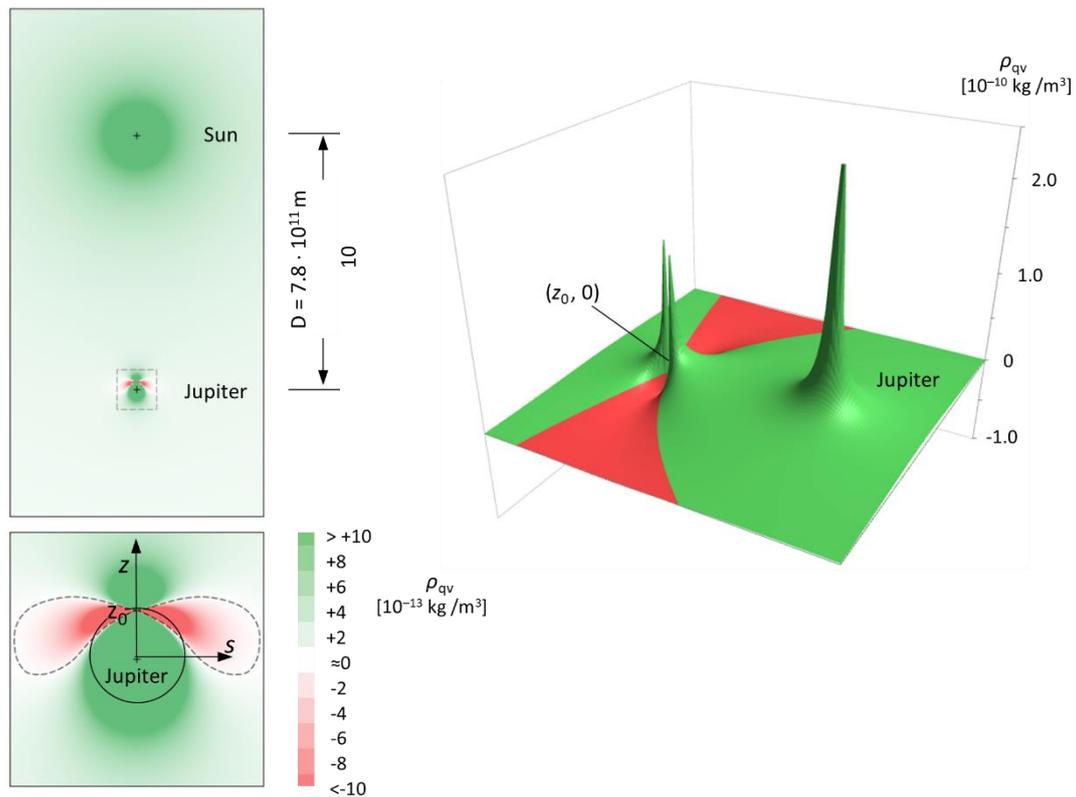

Figure 4. Distribution of gravitational charge density for Jupiter and Sun. Left: 2-dimensional plot where the small, boxed region around Jupiter has been enlarged in the picture below. As in Figure 3 dashed line denotes points with zero density and represents a border between negative densities enclosed by the line and positive densities outside. Right: 3D plot of the region around Jupiter.

7Numerical calculations for different masses $m_1$ and $m_2$ and distances $D \ll R_{sat}$ (where $R_{sat}$ is saturation radius of more massive body) lead to a general picture presented in Figure 3. Note the existence of relatively large red regions with a significant *negative* effective gravitational charge density.

Figure 4 corresponds to the case of Sun and Jupiter; the two-dimensional part of the figure is one more example of plots presented in Fig. 3 with $m_1 \approx m_2/1000$. The corresponding 3-dimensinsal plot shows the granular structure, i.e., the existence of regions in which the effective gravitational charge density is much larger than the average density. This granularity should exist at all scales: Solar System, galaxy, and cluster of galaxies. Hence, it is not surprising that Figure 4 has striking similarity with granular structure revealed for clusters of galaxies (See figures 27, 28 and 29 in Natarjan et al. 2017).

Let us focus on the enlarged square box at the bottom of the left side of Fig.4. The radius of the small circle around Jupiter is $r_c = D/\left(1 + \sqrt{M/m}\right)$ where M and m are respectively the mass of the Sun and planet and D is the distance between them (For Jupiter $D = 7.78 \times 10^{11} m \approx 5.2 AU$). The radius $r_c$ is defined so that on the z axis, at point $(z_0, 0)$ which is at distance $D\sqrt{M/m}$ from the Sun and distance $r_c$ from the Jupiter, Newtonian accelerations caused by the Sun and the Earth cancel each other. For Jupiter $r_c = D/33.38 = 2.4 \times 10^{10} m = 0.16 AU$. The side of the square box containing the whole dashed line (i.e., the whole region with negative density) is a little bit less than $5r_c$. Region presented on the right side of Fig. 4 is even smaller, only a little bit more than distance $r_c$ from Jupiter situated in the centre of the largest peak.

Negative densities within the *dashed* contour span over a few orders of magnitude; in the brightest red region (which is the only region included on the right side of Fig. 4), the maximum of density is a few times larger and minimum a few times smaller than $10^{-12} \, kg/m^3$. Hence region of negative density is not flat as it looks at the plot, but differences are too small to be made visible on a plot that includes a tiny region with peak of positive densities that are larger more than two orders of magnitude.

Note, that numerical calculations revealed the existence of two smaller peaks (see, the right-hand side of Fig. 4) which are in the region of transition from the gravitational field dominated by Sun to the gravitational field dominated by Jupiter.

Let us underscore again the existence of *red region* with negative density; in the case of Jupiter the effective negative gravitational charge of this region is of the order of $10^{20} kg$. If such a negative effective gravitational charge exists in Solar System, it can be revealed by careful study of trajectory of a spacecraft lunched from the Earth and approaching Jupiter through this region.

By the way, the existence of negative gravitational charge density (together with negative and positive over-densities discussed in the next section) suggests a revision of important efforts to reveal if the gravitational polarization of the quantum vacuum is compatible with orbits of planets (Iorio 2019, Pavel and Kroupa 2020).

### 3.2 Case Study: Sun-Saturn

Let us consider Sun and a planet (for instance Saturn) as an isolated two-body system. Of course, this is a rough approximation but helping us to understand complexity of the gravitational polarization of the quantum vacuum in Solar System.

If, in calculations, Sun is considered as the only source of the gravitational polarization of the quantum vacuum (Hajdukovic, 2020a), distribution of the effective gravitational charge density has spherical symmetry, with an *always* positive value $\rho_{qvSun}(r) = 2P_{gmax}/r$ ( r is the distance from the Sun) and a constant acceleration $g_{qvmax} = 4\pi G P_{gmax}$. If a second body (in our case planet Saturn) is included in calculations as a second source of gravitational polarization, this simple pattern of gravitational polarization is replaced by a more complex one. There are three main features of new distribution of the effective gravitational charge density. First, spherical symmetry converts to cylindrical symmetry. Second, the effective gravitational charge density $\rho_{qv}(z,s)$, caused by Sun and Saturn is different from the corresponding density $\rho_{qvSun}(z,s)$ calculated when Sun was considered as the only source of polarization; as presented on Figure 5a, there are regions in which $\rho_{qv}(z,s) > \rho_{qvSun}(z,s)$ and $\rho_{qv}(z,s) < \rho_{qvSun}(z,s)$. Third, and the most fundamental result is that (as shown at Figure 5b) there are regions with a *negative* density $\rho_{qv}(z,s)$.

In the Figure 5a, Sun is denoted by a cross on the left side, and Saturn by a cross near the centre of the figure. This figure, based on numerical calculations, emphases that presence of Saturn, shifts the effective gravitational charge density to values which are less than and greater than the density $2P_{gmax}/r$ calculated with the Sun as the only source of polarization; the corresponding regions are respectively denoted by orange and blue. A darker colour (a darker orange or a darker blue) denotes a larger deviation from density caused by the Sun only. Let us underscore that orange and blue do not present densities but *deviations* (we may also say *over-densities*) from density $\rho_{qvSun}(r)$; that's why we cannot use green and red, denoting densities in the previous figures 3 and 4. It is obvious that positive over-density within the orbit of Saturn produces an additional, positive acceleration of Saturn towards the Sun, while positive over-density outside the orbit causes a negative acceleration. Concerning negative over densities (orange colour), they contribute to a negative acceleration if they are on the right side of Saturn (note that it is dominant region of negative over-densities) and to positive acceleration if they are on the left side of Saturn.

Sun is at nearly 10 Astronomical Units from Saturn. Hence, global plot at Figure 5a corresponds to a very large region (the side of the square is about 50AU) and many details are hidden; especially details in the vicinity of Saturn which are shown in Figure 5b.

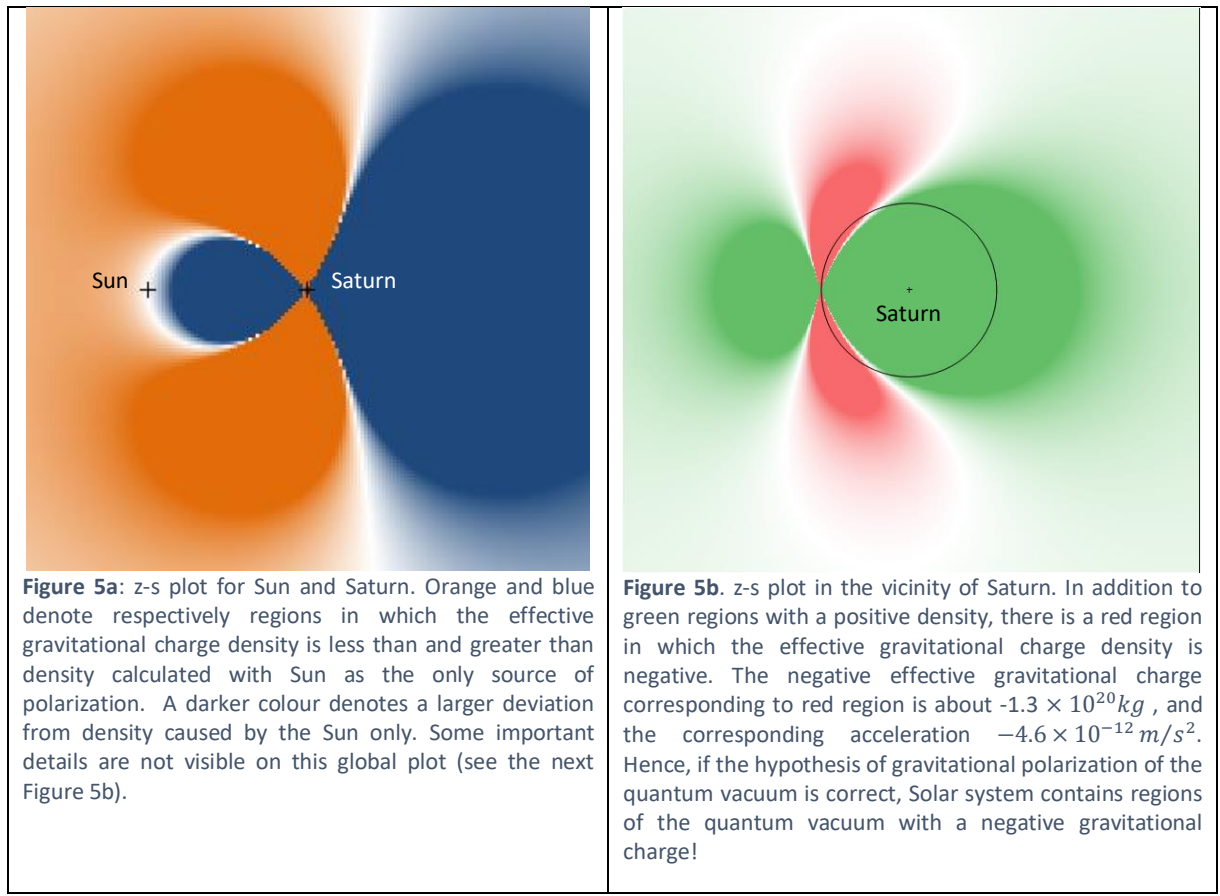

**Figure 5a**: z-s plot for Sun and Saturn. Orange and blue denote respectively regions in which the effective gravitational charge density is less than and greater than density calculated with Sun as the only source of polarization. A darker colour denotes a larger deviation from density caused by the Sun only. Some important details are not visible on this global plot (see the next Figure 5b).

**Figure 5b**. z-s plot in the vicinity of Saturn. In addition to green regions with a positive density, there is a red region in which the effective gravitational charge density is negative. The negative effective gravitational charge corresponding to red region is about $-1.3 \times 10^{20} kg$, and the corresponding acceleration $-4.6 \times 10^{-12}\ m/s^2$. Hence, if the hypothesis of gravitational polarization of the quantum vacuum is correct, Solar system contains regions of the quantum vacuum with a negative gravitational charge!

In Figure 5b, red colour denotes a region in which the effective gravitational charge density is *negative*. The distance of red points from Saturn is mainly between 1/6 to 1/2 of one Astronomical

Unit. The corresponding *negative* effective gravitational charge of this region of the quantum vacuum is $m_{qv}^- = -1.3 \times 10^{20} kg$ (this value is obtained by numerical integration). This is a relatively big mass close in magnitude to mass of significant dwarf planets as for instance (55637) 2002 UX$_{25}$ and 90482 Orcus. Let us say this in more illuminative words. If the gravitational field of Saturn is "switched off", the red region has a positive gravitational charge of about $4.9 \times 10^{19} kg$ . However, if gravitational field of Saturn is "switched on", the effective gravitational charge of the region becomes *negative* and nearly three times larger in magnitude than was gravitational charge calculated neglecting contribution of Saturn to gravitational polarization. The corresponding gravitational acceleration of Saturn caused by this negative charge is $g_{qv}^- = -4.6 \times 10^{-12} \, m/s^2$. Note that this acceleration has a negative sign, i.e., repels Saturn from the Sun with an acceleration which has a magnitude of nearly 10% of $g_{qvmax}$.

The radius of the circle around Saturn (Figure 5b) is $r_c = D/\left(1 + \sqrt{M/m}\right) = 2.38 \times 10^{10}$m, while the side of the square box (that contains the whole region of negative densities) is a little bit less than $6r_c$. This means that the position of the "island" of a negative effective gravitational charge density is accurately known.

Saturn is not an exception; regions with a negative effective gravitational charge density may exist in the vicinity of other planets and celestial bodies. Hence, if hypothesis that quantum vacuum is an "ocean" of virtual gravitational dipoles is correct, Solar system is populated with, let's say so, "minor planets" or "clouds", composed of quantum vacuum with a negative gravitational charge. Let us underscore again that the existence of such a cloud might be eventually revealed by careful study of trajectory of a spacecraft lunched from the Earth and approaching planet through this cloud. A study of the expected deviations from the Newtonian paths remains a major task not performed in the present paper.

Of course, distribution of density and the corresponding numerical results would be modified if one or more additional sources of gravitational polarization are included. On the global plot 5a, the main missing source of the gravitational polarization is Jupiter which orbits around the Sun at a distance that is roughly one-half of the distance of Saturn. On the local plot 5b, the main missing source of the gravitational polarization is Titan, the biggest of satellites of Saturn, nearly as massive as planet Mercury. This suggests that the exact acceleration of Saturn caused by the quantum vacuum, and in general compatibility of the gravitational polarisation of the quantum vacuum with the observed orbits of planets, demands creation of a new generation of ephemerides, with the included quantum vacuum.

## 4. Discussion

We have derived the exact analytical expression for the effective gravitational charge density of the quantum vacuum caused by two point-like bodies. The significance of this result is twofold. First, it is important as the only possible analytical solution for more than one body; for $N \geq 3$ bodies only numerical solutions are possible. Second, this solution gives the necessary fundamental knowledge for the future development of numerical methods and simulations that include quantum vacuum; for instance, it is of major importance to generalize *N-body simulations in order to include the gravitational impact of the polarized quantum vacuum*.

The most fundamental result of our study is that, if the hypothesis of the gravitational polarization of the quantum vacuum is correct, there are regions of the quantum vacuum (already within the Solar System) with a negative gravitational charge density. Such a feature is the exclusive prediction of this theory, not existing in competing theories as MOND or theories that include gravitational repulsion between matter and antimatter (See Hajdukovic 2020b, for a brief review of these theories).

So far, physicists are *not able* to explain the observed gravitational fields in galaxies and clusters of galaxies. The proposed solutions, mainly dark matter and MOND (Modified Newtonian Dynamics) are just speculations. Of course, the gravitational polarization of the quantum vacuum is also a speculation; the appealing feature is that the observed phenomena may be explained without invoking dark matter or modification of gravity. If the hypothesis of the gravitational polarization of the quantum vacuum is correct, in principle, we can determine the gravitational field in a galaxy or cluster of galaxies using only our law of gravity and the known distribution of ordinary matter (i.e., Standard Model matter made of quarks and leptons interacting through the exchange of gauge bosons; in astronomical jargon called baryonic matter). More precisely, the effective gravitational charge density of the quantum vacuum can be numerically calculated from Eq. (3) if we know the Newtonian acceleration $\mathbf{g}_N(x_1, x_2, x_3)$ caused by the distribution of baryonic matter. Numerical methods and simulations are crucial for both, dark matter and MOND paradigm; the quantum vacuum paradigm cannot have a fair chance in competition with other paradigms without the development of an appropriate numerical tool.

**Contribution of authors**: The first author contributed with physical ideas and analytical work, while the second author developed a preliminary program (DOI: 10.6084/m9.figshare.13554686) for numerical calculations and did the basic calculations together with the corresponding figures.

**Acknowledgments:**
We want to say thank you to an anonymous Referee, for excellent comments and suggestions. One of authors (D.S. Hajdukovic) is thankful to Zeina Wakim from Geneva, for assistance during the work on this paper.

**Data availability**

Data and program used in this paper are published online. See, Sergej Walter, Numerical computing of the gravitational polarization of the quantum vacuum by two point-like bodies.
DOI: 10.6084/m9.figshare.13554686.

Appendix A: Calculation of the effective gravitational charge density

Let us start with the effective gravitational charge density $\rho_{qv}(z,s)$ given by Eq. (10) which can be rewritten as:

$$\rho_{qv}(z,s) = -P_{gmax}f_g(z,s)\left[\frac{\partial}{\partial z}\frac{g_{Nz}}{g_N} + \frac{\partial}{\partial s}\frac{g_{Ns}}{g_N} + \frac{1}{s}\frac{g_{Ns}}{g_N}\right] - P_{gmax}\left[\frac{g_{Nz}}{g_N}\frac{\partial}{\partial z}f_g(z,s) + \frac{g_{Ns}}{g_N}\frac{\partial}{\partial s}f_g(z,s)\right]. \quad (A1)$$

According to Equations (7) and (9), the corresponding derivatives of $f_g(z,s)$ are:

$$\frac{\partial}{\partial z}f_g(z,s) = -\frac{G^2}{g_N^{3/2}\sqrt{g_{qv\,max}}}\cdot\left[\tanh\left(A\sqrt{g_N/g_{qv\,max}}\right)\mathrm{sech}^2\left(\sqrt{g_N/g_{qv\,max}}\right)\right.$$
$$\left.+A\tanh\left(\sqrt{g_N/g_{qv\,max}}\right)\mathrm{sech}^2\left(A\sqrt{g_N/g_{qv\,max}}\right)\right]$$
$$\cdot\left\{\frac{m_1^2 z}{(z^2+s^2)^3} - \frac{m_2^2(D-z)}{[(D-z)^2+s^2]^3} + \frac{m_1 m_2/2}{(z^2+s^2)^{5/2}[(D-z)^2+s^2]^{5/2}}\right.$$
$$\left[(D-2z)(z^2+s^2)[(D-z)^2+s^2]\right.$$
$$\left.\left.+3(z^2+s^2-Dz)\left[z[(D-z)^2+s^2] - (D-z)(z^2+s^2)\right]\right]\right\}, \quad (A2)$$

$$\frac{\partial}{\partial s}f_g(z,s) = -\frac{G^2}{g_N^{3/2}\sqrt{g_{qv\,max}}}\cdot\left[\tanh\left(A\sqrt{g_N/g_{qv\,max}}\right)\mathrm{sech}^2\left(\sqrt{g_N/g_{qv\,max}}\right) + \right.$$
$$\left.A\tanh\left(\sqrt{g_N/g_{qv\,max}}\right)\mathrm{sech}^2\left(A\sqrt{g_N/g_{qv\,max}}\right)\right]$$
$$\cdot\left\{\frac{m_1^2 s}{(z^2+s^2)^3} + \frac{m_2^2 s}{[(D-z)^2+s^2]^3} - \frac{m_1 m_2}{2}\frac{2s(z^2+s^2)[(D-z)^2+s^2]}{(z^2+s^2)^{5/2}[(D-z)^2+s^2]^{5/2}}\right.$$
$$\left.+\frac{m_1 m_2}{2}\frac{3s(z^2+s^2-Dz)\left[[(D-z)^2+s^2]+(z^2+s^2)\right]}{(z^2+s^2)^{5/2}[(D-z)^2+s^2]^{5/2}}\right\}. \quad (A3)$$

Similarly, using Equations (8) and (9)

$$\frac{\partial}{\partial z}\left[\frac{g_{Nz}(z,s)}{g_N(z,s)}\right] = -\frac{G}{g_N}\left\{\left[\frac{m_1}{(z^2+s^2)^{3/2}} - \frac{3m_1 z^2}{(z^2+s^2)^{5/2}} + \frac{m_2}{[(D-z)^2+s^2]^{3/2}} - \frac{3m_2(D-z)^2}{[(D-z)^2+s^2]^{5/2}}\right] + G\frac{g_{Nz}}{g_N^2}\left[-\frac{2m_1^2 z}{(z^2+s^2)^3} + \right.\right.$$
$$\left.\left.\frac{2m_2^2(D-z)}{[(D-z)^2+s^2]^3} + \frac{m_1 m_2(2z-D)}{(z^2+s^2)^{3/2}[(D-z)^2+s^2]^{3/2}} + \frac{3m_1 m_2(D-z)(z^2+s^2-Dz)}{(z^2+s^2)^{3/2}[(D-z)^2+s^2]^{5/2}} - \frac{3m_1 m_2 z(z^2+s^2-Dz)}{(z^2+s^2)^{5/2}[(D-z)^2+s^2]^{3/2}}\right]\right\}, \quad (A4)$$

$$\frac{\partial}{\partial s}\left[\frac{g_{Ns}(z,s)}{g_N(z,s)}\right] = -\frac{G}{g_N}\left\{\left[\frac{m_1}{(z^2+s^2)^{3/2}} - \frac{3m_1 s^2}{(z^2+s^2)^{5/2}} + \frac{m_2}{[(D-z)^2+s^2]^{3/2}} - \frac{3m_2 s^2}{[(D-z)^2+s^2]^{5/2}}\right] + G\frac{g_{Ns}}{g_N^2}\left[-\frac{2m_1^2 s}{(z^2+s^2)^3} - \right.\right.$$
$$\left.\left.\frac{2m_2^2 s}{[(D-z)^2+s^2]^3} + \frac{2m_1 m_2 s}{(z^2+s^2)^{3/2}[(D-z)^2+s^2]^{3/2}} - \frac{3m_1 m_2 s(z^2+s^2-Dz)}{(z^2+s^2)^{3/2}[(D-z)^2+s^2]^{5/2}} - \frac{3m_1 m_2 s(z^2+s^2-Dz)}{(z^2+s^2)^{5/2}[(D-z)^2+s^2]^{3/2}}\right]\right\}. \quad (A5)$$

Let us underscore that all calculations are made in Newtonian framework, neglecting the framework of General Relativity. As anonymous Referee noticed that it is reasonable in a week gravitational field when Newtonian theory is an excellent approximation. However, an open question remains, how gravitational polarization works in a strong gravitational field described by General Relativity; this Referee's question cannot be satisfactory answered in this paper.